\documentclass{elsart3p}

\newcommand{\kvec}{{\bf k}}
\newcommand{\rvec}{{\bf r}}
\newcommand{\tauvec}{{\bf \tau}}

\usepackage{graphicx}
\usepackage{amssymb}

\begin{document}

\begin{frontmatter}

\title{Singlet and triplet bipolarons on the triangular lattice}

\author[label1]{J.P.Hague}
\ead{J.P.Hague@lboro.ac.uk}
\corauth{Tel: +44-1509-228208.  Fax: +44-1509-223986}
\author[label2]{P.E.Kornilovitch}
\author[label1]{J.H.Samson}
\author[label1]{A.S.Alexandrov}
\address[label1]{Department of Physics, Loughborough University, Loughborough, LE11 3TU, UK}
\address[label2]{Hewlett-Packard Company, 1000 NE Circle Blvd,
Corvallis, Oregon 97330, USA}
\thanks[epsrc]{We acknowledge support from EPSRC grants EP/C518365/1 and EP/D07777X/1}

\begin{abstract}
We study the Coulomb-Fr\"ohlich model on a triangular lattice, looking
in particular at states with angular momentum. We examine a simplified
model of crab bipolarons with angular momentum by projecting onto the
low energy subspace of the Coulomb-Fr\"ohlich model with large phonon
frequency. Such a projection is consistent with large long-range
electron-phonon coupling and large repulsive Hubbard $U$. Significant
differences are found between the band structure of singlet and
triplet states: The triplet state (which has a flat band) is found to
be significantly heavier than the singlet state (which has mass
similar to the polaron). We test whether the heavier triplet states
persist to lower electron-phonon coupling using continuous time
quantum Monte Carlo (QMC) simulation. The triplet state is both heavier and
larger, demonstrating that the heavier mass is due to quantum
interference effects on the motion. We also find that retardation
effects reduce the differences between singlet and triplet states,
since they reintroduce second order terms in the hopping into the
inverse effective mass. {\bf PUBLISHED AS: Journal of Physics and Chemistry of Solids Volume 69, Issue 12, December 2008, Pages 3304-3306 }
\end{abstract}

\end{frontmatter}

% main text
\section{Introduction}

The possible role of local pairs in the cuprates and the origin of
the pairing mechanism are subject to intense debate in the
superconductivity community. Many experiments now demonstrate that
there is a strong electron-phonon (e-ph) interaction in the cuprates. In
particular, isotope effects show that lattice vibrations do play a
role in determining the specific electronic structure of
high-temperature superconductors \cite{zhao1995a}. However, consensus
has not yet been reached about the effects of lattice
vibrations. Theorists proposing a view of intrinsic repulsion-mediated
superconductivity claim that the isotope effects are simply due to
change of the polaron mass, which are the underlying quasi-particles
for their theories, but that otherwise the e-ph interaction does not
lead to the pairing required for superconductivity
\cite{anderson2004a}. Here, we explore a scenario in which the e-ph
interaction is so strong that pairing is into local bipolarons, which
form a Bose-Einstein Condensate with no electronic resistance
\cite{alexandrov1981a}.

The possibility of bipolaronic superconductors in real materials has
been critisised on the grounds that, in the presence of strong
Coulomb repulsion, the e-ph interaction would have to be huge to form bound pairs. 
Thus, individual polarons
forming the bipolaron would be extremely heavy
\cite{demello1998a}. For strong e-ph coupling on rectangular lattices
with only nearest-neighbour hopping, it can be demonstrated that
bipolaron mass is proportional to polaron mass squared
\cite{hague2007d}, indicating that bipolaronic superconductivity on
such lattices could only be a low temperature
phenomenon. However, a more recent scenario involves the possibility that
e-ph interactions are longer range than the Coulomb repulsion \cite{ale96}, leading
to inter-site pairs which can move freely on lattices formed from
triangular plaquettes (crab bipolarons). Here, we consider the
difference between singlet and triplet bipolarons on the triangular
lattice. Given recent results concerning the possibility of $d$-wave
superconductivity mediated via phonons \cite{hague2006b}, we consider
this a first step towards understanding unconventional pairing on a
quantitative footing using numerically exact QMC computations.

\begin{figure}[t]
\begin{center}
\includegraphics[width=0.37\textwidth]{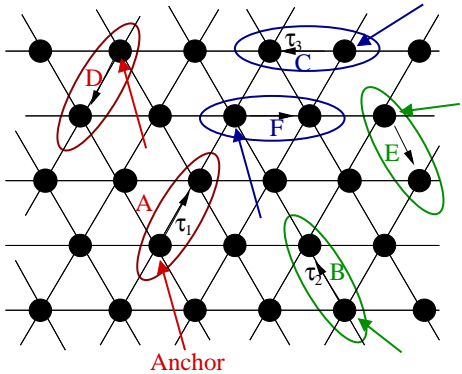}
\end{center}
\caption{(Colour online) Schematic of the small bipolaron on a
triangular lattice with infinite Hubbard repulsion and long range
attraction. For strong e-ph coupling, the energy space is such that
bipolarons must be composed of dimers between nearest neighbour
sites. There are six possible dimer configurations on the triangular
lattice, categorised according to the relative positions of the first
and second inserted particles. Configurations A and D etc.  can be
related through and exchange and translation. The most general 2
particle wavefunction is a linear combination of these six states,
symmetrised according to translational (Bloch) and rotational (angular
momentum) invarience.}
\end{figure}

We study the Coulomb-Fr\"{o}hlich model for long-range e-ph interactions in
quasi-2D materials,
\begin{eqnarray}
H & = & - t \sum_{\langle \mathbf{nn'} \rangle\sigma}
c^{\dagger}_{\mathbf{n'}\sigma} c_{\mathbf{n}\sigma} + \sum_{
\mathbf{nn'}\sigma}V(\mathbf{n},\mathbf{n}') c^{\dagger}_{\mathbf{n}\sigma}
c_{\mathbf{n}\sigma}c^{\dagger}_{\mathbf{n'}\bar{\sigma}} c_{\mathbf{n'}\bar{\sigma}} \nonumber\\ &+&
\sum_{\mathbf{m}} \frac{\hat{P}^{2}_\mathbf{m}}{2M} +
\sum_{\mathbf{m}} \frac{\xi^{2}_{\mathbf{m}} M\omega^2}{2} -
\sum_{\mathbf{n}\mathbf{m}\sigma} f_{\mathbf{m}}(\mathbf{n})
c^{\dagger}_{\mathbf{n}\sigma} c_{\mathbf{n}\sigma} \xi_{\mathbf{m}}\nonumber
\: . \label{eq:four}
\end{eqnarray}
$\xi_\mathbf{m}$ is the ion displacement, sites are numbered by
$\mathbf{n}$ / $\mathbf{m}$ for electrons / ions respectively and $c$
annihilate electrons. The phonons are independent oscillators with
frequency $\omega$ and mass $M$.
$\hat{P}_{\mathbf{m}}=-i\hbar\partial/\partial\xi_{\mathbf{m}}$ is the
ion momentum operator. Since electrons are mobile in the plane,
in-plane Coulomb repulsion $V(\mathbf{n}-\mathbf{n}')$ is heavily
screened. We consider only infinite on site repulsion $U$. The
Fr\"ohlich interaction
is specified via the force function,
$f_{\mathbf{m}}(\mathbf{n})=\kappa\left[(\mathbf{m}-\mathbf{n})^2+1\right]^{-3/2}$,
where $\kappa$ is a constant \cite{alexandrov1999a}. The dimensionless e-ph coupling
$\lambda$ is defined as
$\lambda=\sum_{\mathbf{m}}f^{2}_{\mathbf{m}}(0)/2M\omega^2 zt$ ($z$ is the coordination number). Long-range interactions make polarons light, and less dependent on
lattice type \cite{alexandrov1999a,kornilovitch1998a,kornilovitch1999a,hague2006a}.

\section{Unconventional pairing and the crab bipolaron}

\begin{figure}[t]
\begin{center}
\includegraphics[width=0.25\textwidth]{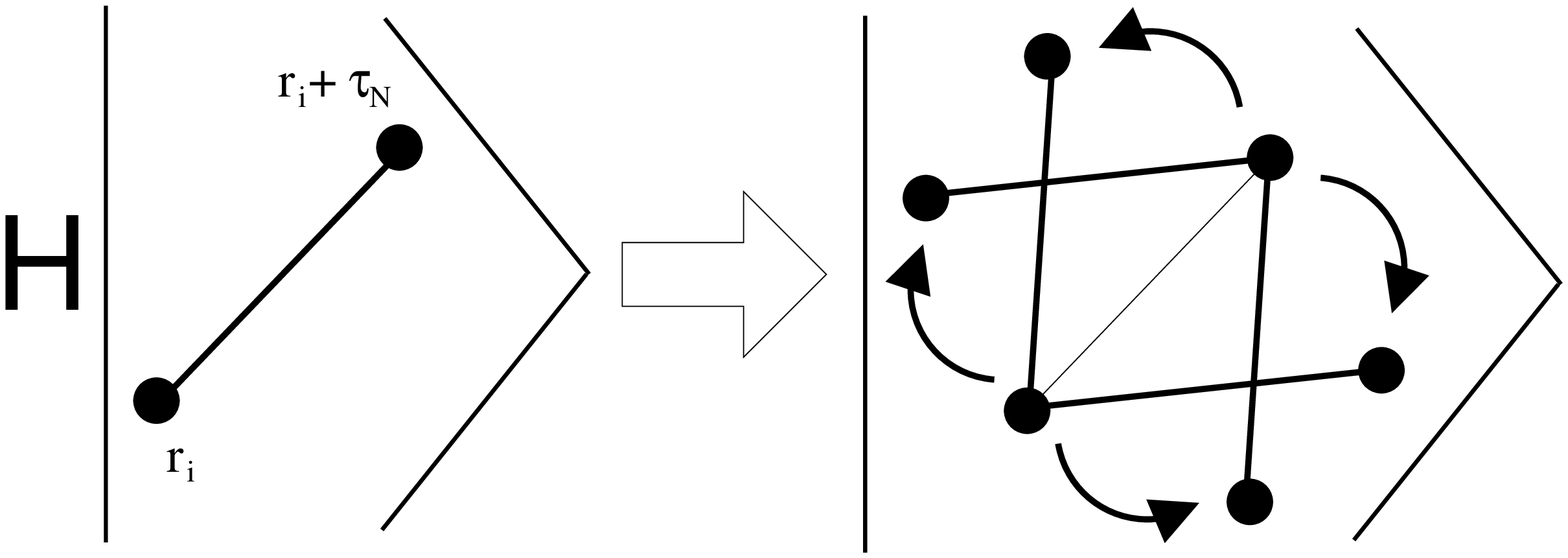}
\end{center}
\caption{Schematic of the hopping term as a set of rotations, showing
that the hopping operator is also related to the angular
momentum operator.}
\end{figure}

To obtain insight into the crab bipolaron, we consider pairs in the
limit where on-site Coulomb repulsion is infinite, inter-site
attraction is large and phonon frequency is large (anti-adiabatic limit $\omega > zt\lambda$). Thus the lowest energy states span a subspace of
near-neighbour pairs. The most general wavefunction in this basis is,
\begin{equation}
|\psi\rangle = \frac{1}{NR}\sum_{\rvec_i} e^{i\kvec.\rvec_i}\sum_{l=1}^{R}e^{\frac{ilm2\pi}{R}}\sum_{n=1}^{R}a_n A^{(n+l)\dagger}_{\rvec_{i}}|0\rangle
\label{eqn:wfgeneral}
\end{equation}
that is that Bloch's theorem and the equivalent for rotated angular
momentum states have been applied to a linear combination of all
neighbour pairs. $R$ is the order of rotational symmetry, $m$ is the
angular momentum. $A^{(N)\dagger}_{\rvec_i} = c^{\dagger}_{\rvec_i}c^{\dagger}_{\rvec_i+\tau_N}$ and $\tau_N$ are the near-neighbour vectors.

\begin{figure}[ht]
\begin{center}
\includegraphics[height=0.4\textwidth,angle=270]{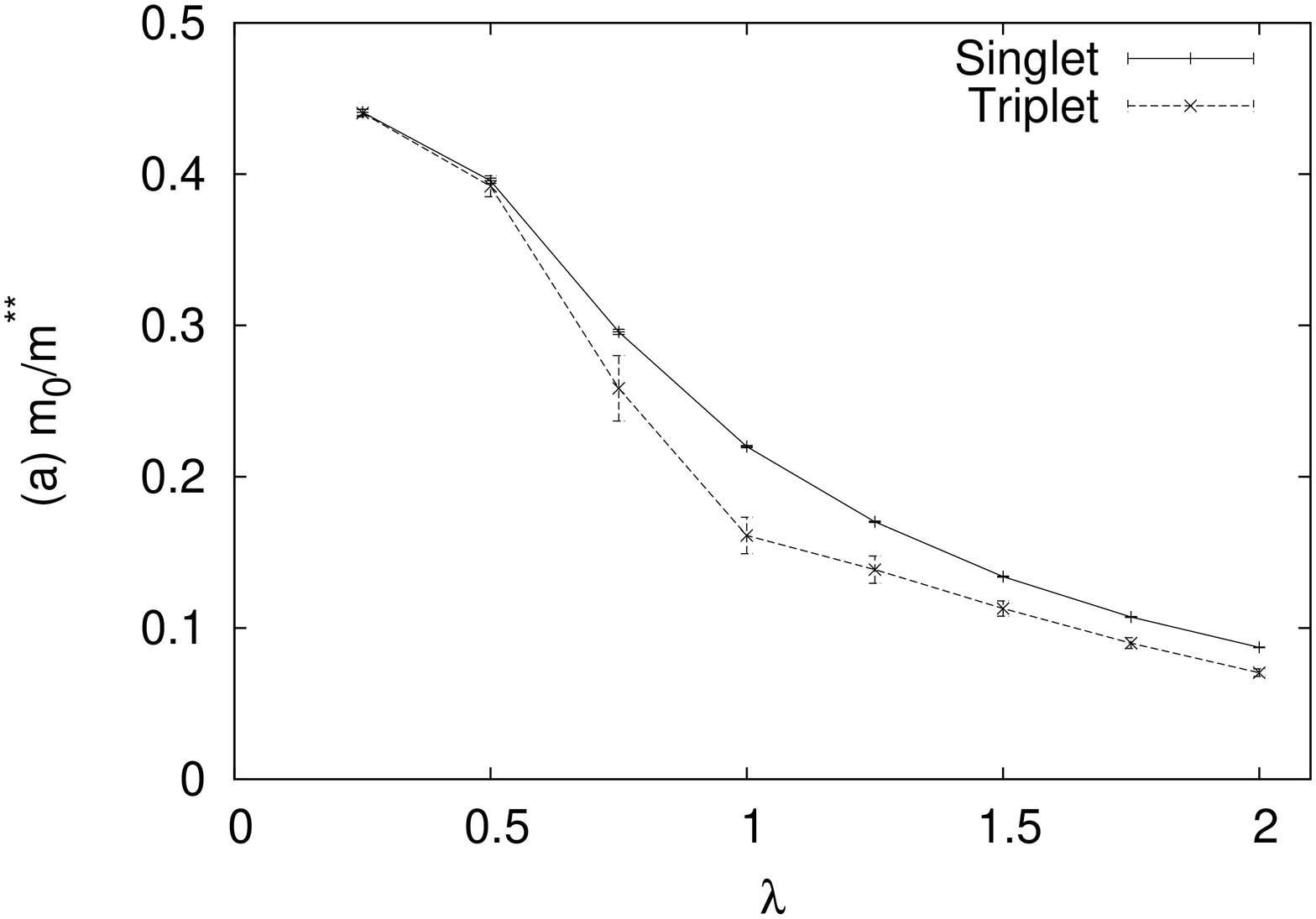}
\includegraphics[height=0.4\textwidth,angle=270]{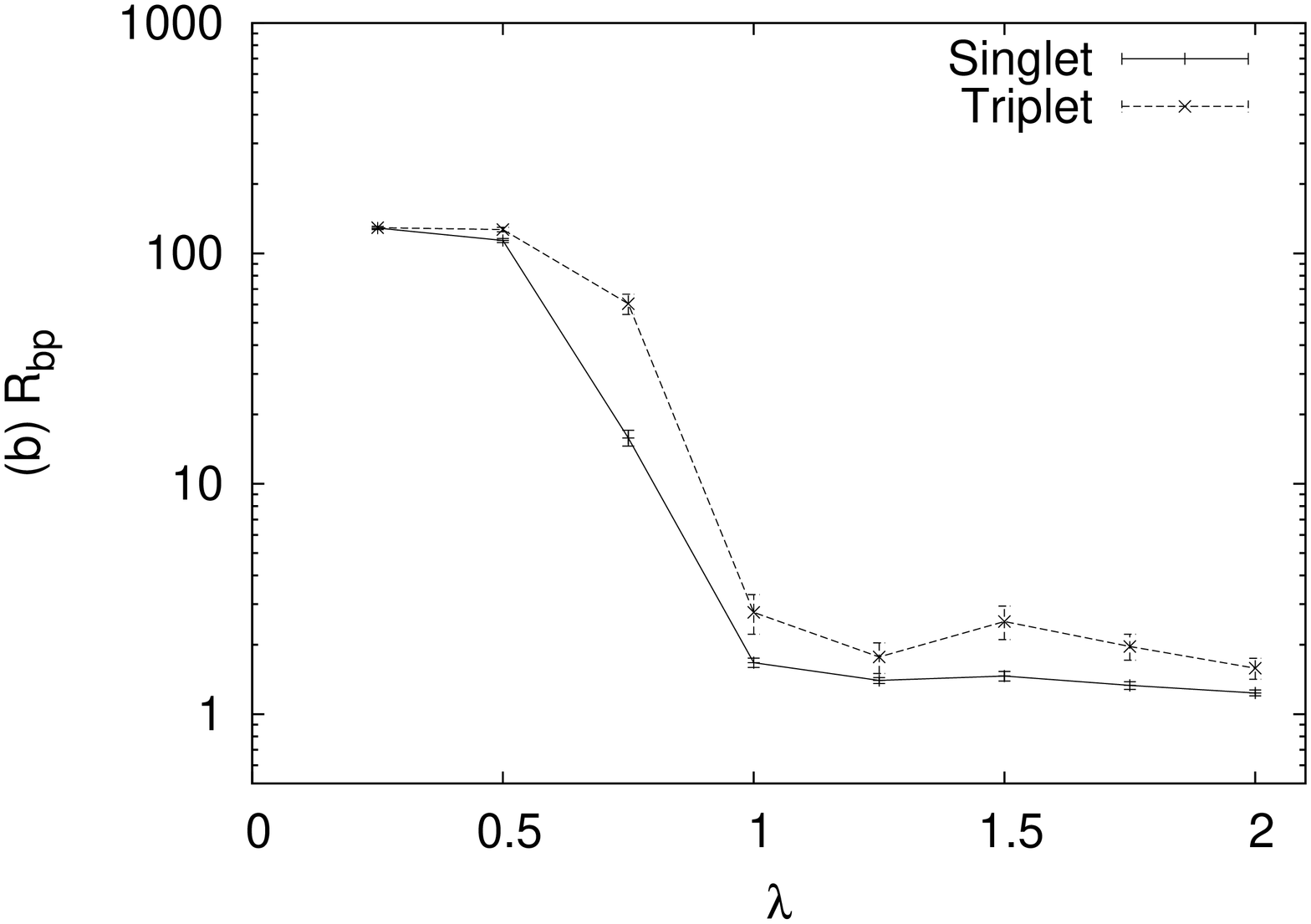}
\end{center}
\caption{\label{fig:figure3} Singlet and triplet states on a
triangular lattice with infinite Coulomb repulsion. $\bar{\beta}=3.5$. Paths are constrained to be within 200 lattice spacings of each other. Measurements were taken every 10 MC steps, with error bars computed by blocking groups of 1000 measurements. Results are
computed for $\bar{\omega}=4$. Panel (a) shows the inverse masses of
singlet and triplet bipolarons, and panel (b) their radii. The masses
of the triplet bipolarons are generally larger than their singlet
counterparts. The radii of triplet bipolarons are also larger, as
should be expected due to the node in the wavefunction. However, this
makes the heavier mass more surprising, since larger (i.e. less well
bound) bipolarons would be expected to be lighter. The apparent
convergence of the singlet and triplet properties at large $\lambda$
is due to retardation effects.}
\end{figure}

There are a number of symmetries that can be used to simplify this
expression: $A^{(N+R/2)\dagger}_{\rvec_i} =
-A^{(N)\dagger}_{\rvec_i-\tau_N}$, $A^{(N+R)\dagger}_{\rvec_i} =
A^{(N)\dagger}_{\rvec_i}$, making the substitution $\bar{a}_n = a_n +
(-1)^ma_{n+R/2}$ and noting that $e^{2\pi i m(l+R/2)/R} = (-1)^m
e^{\frac{ilm2\pi}{R}}$, the
wavefunction can be simplified to:
\begin{eqnarray}
|\psi\rangle & = & \frac{1}{NR}\sum_{\rvec_i} e^{i\kvec.\rvec_i}\sum_{l=1}^{R/2}e^{\frac{ilm2\pi}{R}}\nonumber\\
& & \times\sum_{n=1}^{R/2}\bar{a}_{n-l} (A^{(n)\dagger}_{\rvec_{i}}+(-1)^{m}A^{(n+R/2)\dagger}_{\rvec_{i}})|0\rangle
\label{eqn:wfsimple}
\end{eqnarray}
i.e. the problem is reformulated as $R/2$ singlet and $R/2$ triplet bands.

The effect of the hopping term on pairs can be written as a rotation,
showing that hopping also has some of the properties of an angular
momentum operator (see figure 2).
\begin{eqnarray}
\tilde{H}_{tb} A^{(N)\dagger}_{\rvec_i} = -\tilde{t} & & \left( A^{(N+1)\dagger}_{\rvec_i} + A^{(N-1)\dagger}_{\rvec_i}\right.\\
& & \left. -A^{(N+1+R/2)\dagger}_{\rvec_i+\tauvec_N}-A^{(N-1+R/2)\dagger}_{\rvec_i+\tauvec_N}\right)\nonumber
\end{eqnarray}
If the basis didn't just contain nearest neighbour pairs, then
$\tilde{H}_{tb}$ would introduce $m$ dependent functions into the
result. However for only nearest neighbour pairs,
eq. \ref{eqn:wfsimple} can be rewritten as,
\begin{equation}
|\psi\rangle = \frac{1}{NR}\sum_{\rvec_i} e^{i\kvec.\rvec_i}\sum_{l=1}^{R/2}e^{\frac{ilm2\pi}{R}} \sum_{n=1}^{R/2}\bar{a}_{n-l} \bar{A}^{(n)\dagger}_{\rvec_{i}}|0\rangle
\label{eqn:singtrip}
\end{equation}
where, $\bar{A}^{(n)\dagger}_{\rvec_{i}} = A^{(n)\dagger}_{\rvec_{i}}
+ (-1)^{m}A^{(n+R/2)\dagger}_{\rvec_{i}}$ and $\bar{a}_n = a_n +
(-1)^{m}a_{n+R/2}$. Thus, there is no dependence
on angular momentum. $s$, $d$ etc. states are degenerate singlets, $p$, $f$
etc. are degenerate triplets. Initially, the degeneracy of the angular momentum
states seems surprising. However, since only the near-neighbour
subspace is considered, there are only 6N degrees of freedom. As an
increasing number of possible pairing arrangements is considered (for
example if $U$ were not infinite), the total number of degrees of
freedom would increase, and the degeneracy would be lifted.

As we have previously discussed, singlet and triplet states can be
mapped onto a dimer lattice, with hopping $-\tilde{t}$ for singlets
and $\tilde{t}$ for triplets (the difference in hopping sign can be demonstrated by considering only two particles on a singlet trianglular molecule, symmetrised according to eq. \ref{eqn:wfsimple})  \cite{hague2007b}. Thus one can determine the secular equations:
\begin{equation}
E\langle N|\psi\rangle = \langle N|\tilde{H}|\psi\rangle
\end{equation}
where $|N\rangle = \bar{A}^{(N)\dagger}|0\rangle$. The solution to
those equations is given by the equation,
\begin{equation}
{\rm det}|H_{\kvec} - E| = {\mp \tilde{t}}\left| \begin{array}{ccc}\pm E/\tilde{t} & 1+\gamma_{\kvec}^* & 1+\beta_{\kvec} \\1+\gamma_{\kvec} & \pm E/\tilde{t} & 1+\alpha_{\kvec}^* \\1+\beta_{\kvec}^* & 1+\alpha_{\kvec} & \pm E/\tilde{t}\end{array} \right| = 0 
\end{equation}
with $\alpha_{\kvec} = \exp\left(-ia k_x/2 + i\sqrt{3}a k_y /2)\right)$, $\beta_{\kvec} = \exp\left(-ik_x a/2 -i \sqrt{3} k_y a/2\right)$
and $\gamma_{\kvec} = \exp\left(ik_x a\right)$.

There are 6 bands resulting from the diagonalisation of the
Hamiltonian. Three singlet (corresponding to even $m$) and three
triplet (odd $m$ states),
\begin{eqnarray}
E_1(\kvec) & = & V_{\rm min}\pm\tilde{t}(-1-\sqrt{3+\epsilon(\kvec)/\tilde{t}})\\
E_2(\kvec) & = & V_{\rm min}\pm\tilde{t}(-1+\sqrt{3+\epsilon(\kvec)/\tilde{t}})\\
E_3(\kvec) & = & V_{\rm min}\pm 2\tilde{t}
\end{eqnarray}
where $\epsilon(\kvec) = -2\tilde{t} (\cos k_x a + \cos\left(k_x a/2
-\sqrt{3} k_y a/2\right) + \cos\left(k_x a/2 +\sqrt{3} k_y
a/2\right))$ is the polaron band structure in the antiadiabatic
limit. An unusual result is that the singlet bipolaron mass is
proportional to the polaron mass, $m^{**}_s = 6m^*$ while the triplet
bipolaron mass is infinite. In the rest of this article, we compute
singlet and triplet states computed using a continuous time QMC algorithm to determine which of these attributes survive
into the intermediate coupling and phonon frequency regime
\cite{hague2007a,hague2007b}.

\section{Quantum Monte-Carlo results}

QMC simulations of the crab bipolaron state have been carried out for
singlet states \cite{hague2007a,hague2007b}. We have also simulated
triplet bipolarons on the chain \cite{hague2007c}. It is of interest to
determine if the limiting analytic results discussed in the previous
section can be realised in numerical simulations. We have previously
determined that there are significant differences between the masses
of bipolarons on staggered and rectangular ladders
\cite{hague2007b,hague2007d}. We have also found that masses of
polarons and bipolarons on lattices with triangular components have
similar magnitude \cite{hague2007a}.

We compute numerical results from the continuous
time QMC algorithm for a bipolaron on a triangular
lattice. The singlet and triplet properties are computed by
symmetrising the waveunctions. So the triplet bipolaron
picks up a sign change on exchange, whereas the singlet always has
positive sign. This affects the estimators, which depend on the sign
as,
\begin{equation}
R_{bp}=\left\langle s \sqrt{\frac{1}{\beta}\int_{0}^{\beta}(\mathbf{r}_{1}(\tau)-\mathbf{r}_2(\tau))^2
d\tau}\right\rangle
\end{equation}
\begin{equation}
\frac{m_0}{m_i^{**}} = \lim_{\beta\rightarrow\infty} \frac{1}{\bar{\beta}}\left\langle s\Delta {\bf r}^2_i \right\rangle
\end{equation}
For configurations where the end points of the paths are together, $s=0$. $\Delta {\bf r}$ is defined as in ref. \cite{kornilovitch1998a}.

Figure \ref{fig:figure3} shows the effective mass and size of singlet
and triplet states on a triangular lattice with infinite local Coulomb
repulsion. Results are computed for $\omega/t=4$. Panel (a) shows
the inverse masses of singlet and triplet bipolarons, and panel (b)
their radii. It can be seen that the masses of the triplet bipolarons
are generally larger than their singlet counterparts. The radii of
triplet bipolarons are also larger, as should be expected due to the
node in the wavefunction. However, this makes the heavier mass more
surprising, since larger bipolarons would be expected to be lighter
since they are less well bound. The larger mass is thus due to quantum
interference effects in the motion. This can be best understood by imagining the tight binding model constructed for hopping of bipolarons. The bipolaron overlap integrals giving the effective hopping are highly sensitive to the form of the binding wavefunction, and it is clear that the node in the triplet wavefunction will make those integrals (i.e. the effective hopping) much smaller leading to smaller effective mass. For larger couplings, $zt\lambda >
\omega$, the antiadiabatic approximation used in the previous section
does not hold. Thus the largest differences between triplet and
singlet states are found around $\lambda=1$.

The apparent convergence of singlet and triplet states at large
coupling is initially surprising. However, analysis of the sign can
shed light on this effect. For weak coupling, any bipolarons are
weakly bound, so singlet-triplet splitting must be very small, since
it can't be larger than the singlet binding energy. Since $\Delta
E_{st}=-(\ln\langle s \rangle)/\beta$, the average sign must be
approximately 1. At large coupling, paths must be nearly straight at
very strong coupling due to retardation effects. Since there is
infinite Coulomb repulsion, at least two kink insertions are required
on each path for an exchange, exchanges become increasingly rare, and
the sign must return to 1. So the total sign dips (and
thus the difference in mass is largest) when
$zt\lambda\sim\omega$.

\section{Conclusion}

We have computed properties of bipolarons with angular momentum on
triangular lattices. The mass of the singlet bipolaron is of the order
of the polaron mass for a wide range of couplings. However, we find
that triplet bipolarons have much larger mass for coupling
$zt\lambda\sim\omega$. A surprising result here is that retardation
effects at very large $\lambda$ reduce the difference between singlet
and triplet properties when the local Coulomb repulsion is
infinite. Understanding of the full range of bipolaron behaviours, especially on
unconventional lattices is clearly important,
and will form the subject of future publications.

\bibliographystyle{elsart-num}
\bibliography{triangular_unconventional}

\end{document}